\documentclass[
  nojss]{jss}

\usepackage{orcidlink,thumbpdf,lmodern}

\usepackage[utf8]{inputenc}

\author{
Victor Freguglia~\orcidlink{0000-0002-6189-4453}\\University of
Campinas \And Thiago
Paulichen~\orcidlink{0009-0002-5505-2409}\\University of Campinas
}
\title{\pkg{bacontrees}: An extensible framework for Bayesian Context
Trees in R}

\Plainauthor{Victor Freguglia, Thiago Paulichen}
\Plaintitle{bacontrees: An extensible framework for Bayesian Context
Trees in R}
\Shorttitle{\pkg{bacontrees}: Bayesian Context Trees in R}

\Abstract{
Context tree models, or variable-length Markov chains, describe symbolic
sequences whose dependence structure varies with context length.
Bayesian formulations of these models enable uncertainty quantification
and regularization through priors on tree structures and transition
probabilities. Despite their theoretical appeal, practical
implementations have been fragmented and model-specific. The
\pkg{bacontrees} package for \proglang{R} provides a unified framework
for Bayesian Context Trees by defining core data structures and a
consistent interface for building, modifying and manipulating trees.
Rather than implementing a single inference method, \pkg{bacontrees}
offers the computational foundation upon which diverse Bayesian and
frequentist algorithms can be developed. This separation of
representation and inference promotes reproducibility and facilitates
experimentation with new modeling strategies. The package aims to
support applications in information theory, linguistics, and other
domains where context-based probabilistic modeling is central.
}

\Plainkeywords{Bayesian inference, context trees, variable-length Markov
chains, statistical software, R package, sequential data modeling}

\Address{
    Victor Freguglia\\
    University of Campinas\\
    Institute of Mathematics, Statistics and Scientific Computation\\
  E-mail: \email{victorfs@unicamp.br}\\
  
    }

\usepackage{dsfont, amsmath, amsthm, amsfonts} \usepackage[]{forest} \usepackage{booktabs}

\begin{document}

\newcommand{\cA}{\mathcal{A}}
\newcommand{\cT}{\mathcal{T}}
\newcommand{\Prb}{\mathds{P}}
\newcommand{\cZ}{\mathcal{Z}}
\newcommand{\cz}{\mathfrak{z}}
\newcommand{\taua}{\tau_a}
\newcommand{\taub}{\tau_b}
\newcommand{\cI}{\mathcal{I}}
\newcommand{\bZ}{\mathbf{Z}}
\newcommand{\bz}{\mathbf{z}}
\newcommand{\bp}{\mathbf{p}}
\newcommand{\bs}{\mathbf{s}}
\newtheorem{theorem}{Theorem}
\newtheorem{lemma}{Lemma}

\section{Introduction}

Context Trees play a central role in modeling discrete-valued stochastic
processes, providing parsimonious representations of dependence
structures while naturally accommodating long-range dependencies through
variable-length memory. Although estimating transition probabilities for
a given Context Tree is relatively straightforward, inferring the tree
itself is a challenging model selection problem due to the combinatorial
growth of the space of candidate structures.

Over the past decades, numerous algorithms have been proposed to address
this problem. Representative examples include the frequentist Context
Algorithm \citep{buhlmann1999variable}, Bayesian approaches such as the
Bayesian Context Tree (BCT) algorithm \citep{kontoyiannis2022bayesian},
and weighting schemes such as the Context Tree Weighting (CTW) algorithm
\citep{willems2002context}. Despite their statistical differences, these
methods share a common computational foundation: they repeatedly
manipulate context tree structures, traverse nodes, update node-specific
statistics, evaluate likelihoods, and explore different tree
configurations.

From a software perspective, however, these common computational
abstractions have received considerably less attention than the
statistical methodologies themselves. Existing implementations are
typically designed around a specific inference procedure rather than
around the underlying tree representation, making them difficult to
extend or reuse in new settings. For instance, the \pkg{BCT} package
\citep{pkgbct} implements Bayesian inference under the framework of
\citet{kontoyiannis2022bayesian}, while the \pkg{VLMC} package
\citep{pkgVLMC} focuses on frequentist estimation through the Context
Algorithm. Although both packages provide efficient implementations of
their respective methods, they do not expose a general computational
infrastructure for representing and manipulating context trees that can
be readily reused by other algorithms.

Developing new methods for context trees therefore requires much more
than implementing the statistical procedure itself. A considerable
portion of the implementation effort is devoted to maintaining the tree
structure, managing parent-child relationships, activating and pruning
subtrees, storing node-specific statistics, and efficiently updating
quantities that are shared across multiple candidate trees. These
low-level operations are largely common across inference procedures, yet
they are often reimplemented independently for each new algorithm,
increasing development effort and making software harder to maintain and
extend.

General-purpose packages for trees and graphs, such as \pkg{data.tree}
\citep{pkgdata.tree}, \pkg{igraph} \citep{pkgigraph}, and
\pkg{tidygraph} \citep{pkgtidygraph}, provide powerful abstractions for
generic hierarchical structures. However, they are intentionally
independent of the statistical properties that characterize context
trees, such as context activation, node counts, likelihood calculations,
and tree-specific operations commonly required by inference algorithms.
Consequently, although these packages provide useful low-level data
structures, substantial additional development is still required before
they can support statistical methods based on context trees.

The \pkg{bacontrees} package addresses this gap by providing a reusable
computational framework for Context Trees in \proglang{R}. Rather than
implementing a single inference method, the package provides
object-oriented abstractions for representing, manipulating, and
traversing context trees, together with reusable implementations of
common operations required by both frequentist and Bayesian algorithms.
This architecture allows new methods to be developed by building upon a
common infrastructure instead of reimplementing fundamental data
structures and tree manipulation routines. The package also includes
reference implementations of several commonly used procedures,
illustrating how diverse algorithms can naturally share the same
computational foundation.

The name \pkg{bacontrees} derives from \textbf{Ba}yesian
\textbf{Con}text \textbf{Trees}, reflecting the Bayesian motivation
behind the package. While the initial development was motivated by
Bayesian inference, the underlying software architecture was designed to
support a much broader class of algorithms involving context trees. The
package is available on CRAN and can be installed directly using

\begin{CodeChunk}
\begin{CodeInput}
R> install.packages("bacontrees")
\end{CodeInput}
\end{CodeChunk}

This manuscript is organized as follows. \autoref{sec:background}
reviews the statistical background required for the remainder of the
paper. \autoref{sec:architecture} presents the software architecture of
\pkg{bacontrees}, describing its main classes, data structures, and
computational abstractions. \autoref{sec:examples} illustrates the
package through practical examples, and \autoref{sec:features} surveys
its built-in features, datasets, and visualization capabilities.
\autoref{sec:comparisons} compares \pkg{bacontrees} with existing
software for context trees, and \autoref{sec:conclusions} concludes.

\section{Statistical Background}\label{sec:background}

\subsection{Context Trees}\label{context-trees}

Consider an alphabet of length \(m\), denoted by \(\mathcal{A}\), a set
of \(m\) symbols representing the possible values in a discrete
sequence. We call a \textbf{context tree} a rooted \(m\)-ary tree
structure, denoted by \(\tau\), with the following two properties:

\begin{enumerate}
  \item Every node of $\tau$ has either $0$ or $m$ children.
  \item Each set of $m$ sibling nodes with a common parent have nodes labeled
  with each symbol in $\mathcal{A}$.
\end{enumerate}

Context trees can be uniquely represented by the set of paths connecting
the root to each of its leaves. We call these paths \textbf{contexts}
and will use the notation \(s \in \tau\) to represent a context in
\(\tau\). The root node is assigned a special symbol (\(*\)) for a more
clear notation. We also refer to the number of non-root symbols in a
context as the \textbf{depth} of the context, denoted \(d(s)\), and
\textbf{depth of the context tree}, denoted \(d(\tau)\), is the maximum
depth among its contexts \(d(\tau) = \max_{s \in \tau} d(s)\).

A direct implication of the two properties required for the context tree
is that for any string of \(D\) symbols in \(\mathcal{A}\), denoted
\(\mathbf{z}_{-D}^{-1} = (z_{-D}, z_{-D+1}, \dots, z_{-1})\), if
\(d(\tau) \leq D\), there is a context \(s \in \tau\) such that \(s\)
has the form \[(*.z_{-1}.z_{-2}.\cdots.z_{-d(s)}).\] In other words,
there is a \textbf{suffix} of the sequence \(\mathbf{z}_{-D}^{-1}\) such
that the symbols in such suffix, read starting from the highest index,
\(z_{-1}\), to the lowest, match the path from the root to a leaf of the
context tree, i.e., a context. We define a context function, denoted
\(\eta_\tau\), as a map
\[ \eta_\tau : \bigcup_{D \geq d(\tau)}\mathcal{A}^D \rightarrow \tau,\]
such that \(\eta_\tau(\mathbf{z}_{-D}^{-1})\) is the unique context
\(s \in \tau\) that is a suffix of \(\mathbf{z}_{-D}^{-1}\).

\begin{figure}[ht]
  \centering
  \begin{minipage}{0.45\textwidth}
  \centering
  \begin{forest}
  for tree={
    grow=east, parent anchor=east, child anchor=west, edge={draw},
    if n children=0{
      font=\itshape,
      tier=terminal,
      draw,
      edge=dashed
    }{},}
  [*
    [a
      [a [*.a.a]]
      [b [a [*.a.b.a]][b[*.a.b.b]][c[*.a.b.c]]]
      [c[*.a.c]]
    ]
    [b[*.b]]
    [c[*.c]]
  ]
  \end{forest}
  \end{minipage} \hfill
  \begin{minipage}{0.45\textwidth}
  \centering
  \begin{forest}
  for tree={
    grow=east, parent anchor=east, child anchor=west, edge={draw},
    if n children=0{
      font=\itshape,
      tier=terminal,
      draw,
      edge=dashed
    }{},}
    [*
      [0 [*.0]]
      [1
        [0 [*.1.0]]
        [1
          [0
            [0 [*.1.1.0.0]]
            [1 [*.1.1.0.1]]
          ]
          [1 [*.1.1.1]]
        ]
      ]
    ]
  \end{forest}
  \end{minipage}
  \caption{Two examples of Context Trees with alphabet $\mathcal{A}= \{a,b,c\}$ and
  7 contexts (left) and $\mathcal{A}= \{0,1\}$ and 5 contexts (right).
  \label{fig:tree_examples}}
\end{figure}
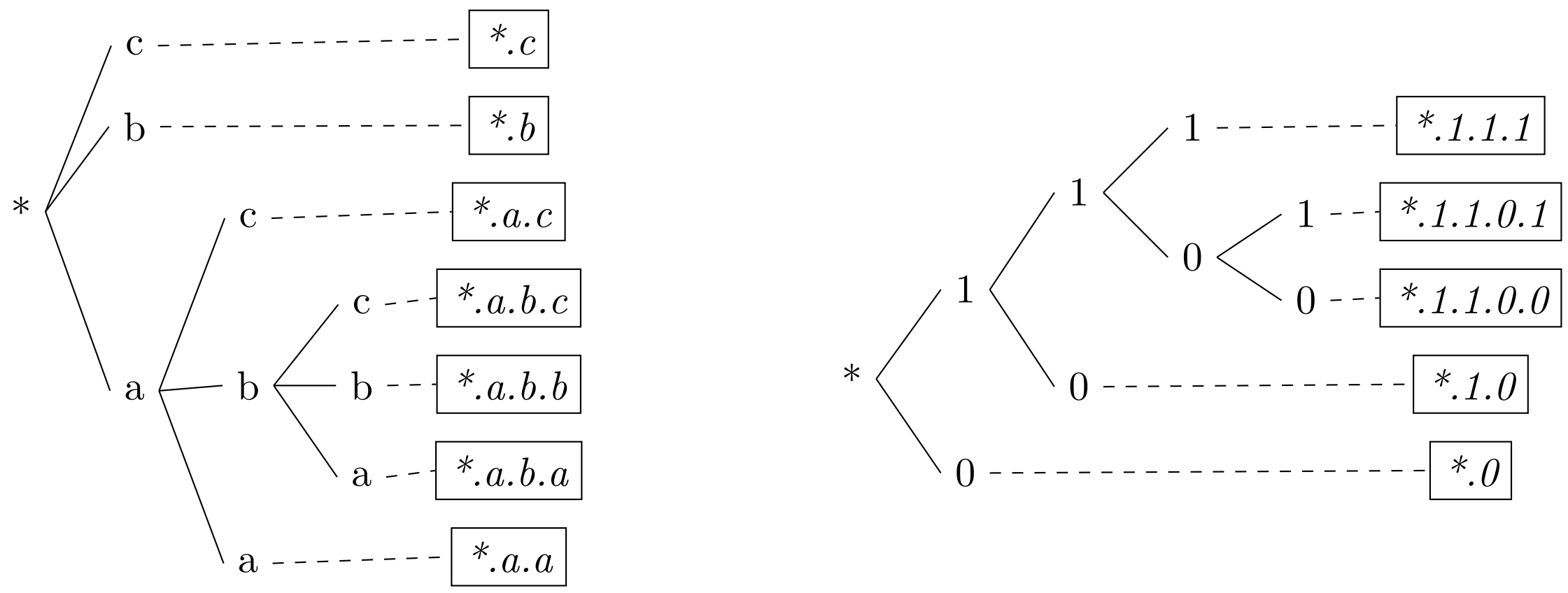

As an example, the right-side tree from \autoref{fig:tree_examples} is
represented by the set of contexts
\(\tau = \{ (*.0), (*.1.0), (*.1.1.0.0), (*.1.1.0.1), (*.1.1.1)\}\) and
it has a depth \(d(\tau) = 4\). Under this context tree, the string
\((0,1,0,1)\) would be mapped to the context \((*.1.0)\) by
\(\eta_\tau\).

\subsection{Variable-Length Markov
Chains}\label{variable-length-markov-chains}

Given a context tree \(\tau\), we introduce the vectors of transition
probabilities
\[\mathbf{p} = (p_{s,k}), {s \in \tau, k \in \mathcal{A}},\] with
\(0 \leq p_{s,k} \leq 1\) for all \(s\) and \(k\), and
\(\sum_{k \in \mathcal{A}} p_{sk} = 1\) for each context \(s\). We also
use the notation \(\mathbf{p}_s\) to represent the entries related to a
context \(s\), \(\mathbf{p}_s = (p_{s,k}), k \in \mathcal{A}\). A
sequence of random variables \(\mathbf{Z}_{1}^{n}\) assuming values in
\(\mathcal{A}\) is a Variable-Length Markov Chain, with respect to
\(\tau\), if the conditional probabilities can be expressed as
\begin{equation}
  \mathbb{P}(Z_t = k \mid \mathbf{Z}_{1}^{t-1} = \mathbf{z}_{1}^{t-1}) =
  \mathbb{P}(Z_t = k \mid \eta_\tau(\mathbf{Z}_{1}^{t-1}) = \eta_\tau(\mathbf{z}_{1}^{t-1})) =
  p_{\eta_\tau(\mathbf{z}_{1}^{t-1}),k}, \text{ for } t \geq d(\tau),
\end{equation} for some vectors \(\mathbf{p}\). In other words, the
conditional probability of the next symbol given the entire history,
depends only on the suffix of the history mapped by the context tree,
and the transition probability depends only on which is the mapped
context.

Throughout this work, we will consider the first \(d(\tau)\) symbols as
fixed values under VLMC models to avoid introducing additional
complexity modeling the first terms. Therefore, given a pair
\((\tau, \mathbf{p})\), the probability of observing a sequence
\(\mathbf{z}_{1}^{n}\) is \begin{equation}
  \mathbb{P}(\mathbf{Z}_{1}^{n} = \mathbf{z}_{1}^{n}\mid \tau, \mathbf{p}) =
  \prod_{t = d(\tau)}^n \prod_{k \in \mathcal{A}} \prod_{s \in \tau} \left(p_{s,k}\right)^{\mathds{1}(z_t = k)\mathds{1}(\eta_\tau(\mathbf{z}_{1}^{t-1}) = s)}.
\end{equation}

Additionally, we will consider that datasets can be composed by a set of
multiple independent sequences, with varying number of observations, but
generated from the same model specified by \((\tau, \mathbf{p})\). These
sets of random sequences will be denoted \begin{equation*}
\mathcal{Z} = \{ \mathbf{Z}^{(i)}, i = 1, \dots, I \}, \text{ with }\mathbf{Z}^{(i)} = {\mathbf{Z}^{(i)}}_1^{n_i},
\end{equation*} Under these definitions, the joint probability of a
specific realization, \(\mathfrak{z}\), of the random set of sequences
\(\mathcal{Z}\) can be expressed as: \begin{equation}
  \mathbb{P}(\mathcal{Z} = \mathfrak{z} \mid \tau, \mathbf{p}) = 
  \prod_{s \in \tau} \prod_{k \in \mathcal{A}} \left(p_{s,k} \right)^{n_{s,k}(\mathfrak{z})},
  \label{eq:prob}
\end{equation} where
\(n_{s,k} = \sum_{i = 1}^I \sum_{t = d(\tau)}^{n_i} \mathds{1}\left( \eta_{\tau}({\mathbf{z}^{(i)}}_{1}^{t}) = s \right) \mathds{1}\left(z_t^{(i)} = k \right)\)
represents the total count of transitions from context \(s\) to state
\(k\) in all sequences.

Therefore, for a fixed context tree, \(\tau\), we can compute the
Maximum Likelihood Estimator of \(\mathbf{p}\), denoted
\(\hat{\mathbf{p}}\), as \begin{equation}
  \hat{p}_{s,k} = \frac{n_{s,k}(\mathfrak{z})}{\sum_{k' \in \mathcal{A}} n_{s,k'}(\mathfrak{z})},
  \label{eq:MLE}
\end{equation} i.e., the proportion of observed transitions from each
context \(s\) to each state \(k\).

\subsection{Bayesian Approach to VLMC}\label{bayesian-approach-to-vlmc}

From a Bayesian perspective, we can consider the context tree to be a
random object with a prior distribution \(\pi(\tau)\) and, in
particular, we consider this distribution to be the product of a weight
function \(w\) evaluated on each context of \(\tau\), i.e.,
\(\pi(\tau) \propto \prod_{s \in \tau} w(s)\). All prior distributions
commonly used in Bayesian Context Trees fit in this definition, although
they may be even simpler, such as the ones from
\citet{willems2002context} and \citet{kontoyiannis2022bayesian}.

Given \(\tau\), we consider \(\mathbf{p}\) to be a random vector
composed by \(|\tau|\) independent \(m\)-dimensional sub-vectors, each
on the simplex space \(\Delta_m\), and we consider them to be
distributed as symmetrical Dirichlet distributions with a
hyper-parameter \(\alpha > 0\). The value \(\alpha = 1/2\) corresponds
to the widely used Krichevsky--Trofimov estimator
\citep{krichevsky1981performance}.

These definitions result in a Bayesian system described as follows:
\begin{align} \label{eq:bayesian_system}
    \tau &\sim \pi(\tau) \propto \prod_{s \in \tau} w(s), & \tau \in \mathcal{T}  \notag \\
    \mathbf{p} \mid \tau &\sim \prod_{s \in \tau} \frac{\Gamma(m\alpha)}{\Gamma(\alpha)^m} \prod_{k \in \mathcal{A}} p_{s,k}^{\alpha - 1} & \mathbf{p} \in \Delta^{|\tau|}_m \\
    \mathcal{Z} \mid (\tau, \mathbf{p}) &\sim \mathbb{P}(\mathcal{Z} = \mathfrak{z}) \text{ as in \eqref{eq:prob}}, & \mathfrak{z} \in \mathcal{A}^{n_1 \times \dots \times n_I} \notag 
\end{align}

An important property of the system described in
\eqref{eq:bayesian_system} is that the independent Dirichlet
distributions are conjugate to the likelihood function. This conjugacy
allows us to integrate the joint probability of the system with respect
to \(\mathbf{p}\). As a result, the marginal posterior distribution
\(\pi(\tau \mid \mathfrak{z})\) can be written, up to a proportionality
constant, as \begin{equation}\label{eq:marginal_posterior}
  \pi(\tau \mid \mathfrak{z}) \propto
  \prod_{s \in \tau} w(s)
  \left(\frac{\Gamma(m\alpha)}{\Gamma(\alpha)^m}\right)
  \frac{\prod_{k = 0}^{m-1}\Gamma(\alpha + n_{s,k}(\mathfrak{z}))}{\Gamma(m\alpha + \sum_{k = 0}^{m-1} n_{s,k}(\mathfrak{z}))} = 
  \prod_{s \in \tau} w(s) q_{\alpha}(\mathfrak{z}).
\end{equation} The Marginal Tree Posterior in
\eqref{eq:marginal_posterior} is a key component of many Bayesian
algorithms, including posterior sampling via Metropolis-Hastings and
computation of Partial Bayes Factors \citep{freguglia2023detecting}.

Having introduced the statistical concepts required to define context
trees and their Bayesian formulation, we now describe the software
architecture that supports their manipulation.

\section{Software Architecture}\label{sec:architecture}

The development of algorithms for Context Trees presents a recurring
software engineering challenge. Although inference procedures differ
considerably from a statistical perspective, they typically rely on the
same underlying computational tasks: representing tree structures,
traversing nodes, maintaining node-specific statistics, and modifying
tree configurations during the execution of the algorithm. As a result,
a substantial portion of the implementation effort is independent of the
statistical methodology itself.

The design of \pkg{bacontrees} is based on the observation that these
common tasks can be isolated into a small set of reusable abstractions.
Rather than implementing a particular inference algorithm, the package
provides a computational representation of context trees together with
the operations required to manipulate them. Statistical procedures are
then implemented by combining these building blocks instead of managing
the underlying tree structure directly.

The architecture follows three hierarchical levels. At the lowest level,
individual nodes are represented by \code{TreeNode} objects, which store
the information associated with a single context and maintain the
relationships with their parent and children. These nodes are organized
into a \code{ContextTree}, which represents the maximal tree associated
with the observed data and provides methods for navigation and
structural manipulation. Finally, the \code{baConTree} class extends
this representation with the additional quantities required for Bayesian
inference, including prior specifications, posterior probabilities, and
cached quantities used by inference algorithms.

An important design principle is the distinction between the maximal
tree and the active tree. The maximal tree stores every context that can
be represented given the observed data and the maximum depth considered.
Individual Context Trees are then represented by selecting a subset of
these nodes as active, rather than by constructing independent tree
objects. Consequently, different candidate trees share the same
underlying data structure and differ only in their activation state.

This representation has important computational consequences. Since all
node-specific information is stored only once, statistics do not need to
be recomputed when exploring different tree configurations. Tree
modifications become local operations that activate or deactivate nodes
while preserving the underlying structure. This design considerably
simplifies the implementation of algorithms that repeatedly traverse or
modify Context Trees, including pruning procedures, posterior
exploration, and model averaging methods.

The following sections describe each component of this architecture in
detail, beginning with the representation of individual nodes and
progressing toward the higher-level abstractions used by inference
algorithms.

\subsection{TreeNode}\label{treenode}

The \code{TreeNode} class is the fundamental building block of the
\pkg{bacontrees} package. Although users rarely manipulate
\code{TreeNode} objects directly, they constitute the basic
computational unit upon which all higher-level classes are built. Each
object represents a single node of a Context Tree and encapsulates both
the structural information required to navigate the tree and the
statistical quantities associated with that context. Since most
algorithms on Context Trees perform computations locally at individual
nodes, this representation provides a reusable abstraction upon which
different inference procedures can be built.

Objects of the \code{TreeNode} class are uniquely identified by their
\emph{path} (\(s\), in the notation from \eqref{eq:bayesian_system}), a
sequence of symbols that defines their position in the tree, and their
\emph{depth} attribute (\(d(s)\)) is automatically derived from the
path. The class includes fields for storing node-specific transition
counts \(n_{s,k}\) (\code{counts}), a logical flag indicating whether
the node is terminal, and a flexible modifiable list (\code{extra}) used
to store auxiliary information required by higher-level algorithms.

From a structural perspective, each node maintains references to its
parent and children, allowing efficient traversal of the tree in both
directions. A variety of methods are provided for managing the node's
state and relationships within the tree, including accessors for
structural attributes (\code{getPath()}, \code{getDepth()},
\code{getParentPath()}, and \code{getChildrenPaths()}). The method
\code{isActive()} indicates whether the node belongs to the current
active tree, which is particularly useful for algorithms that
iteratively modify the tree structure. This lightweight design allows
nodes to be manipulated independently while remaining consistent with
the global tree representation managed by the \code{ContextTree} class.

\code{TreeNode} objects naturally live within a \code{ContextTree},
where they are stored in the \code{nodes} attribute as a named list
indexed by the corresponding contexts. The example below illustrates how
individual nodes can be accessed and queried.

\begin{CodeChunk}
\begin{CodeInput}
R> ct_example2 <- ContextTree$new(dataset = sample(0:1, 100, TRUE), maximalDepth = 3)
R> node <- ct_example2$nodes[["*.1.0"]]
R> node$getPath()      # Returns the path of the node
\end{CodeInput}
\begin{CodeOutput}
[1] "*.1.0"
\end{CodeOutput}
\begin{CodeInput}
R> node$getParentPath()  # Returns the path of the parent node
\end{CodeInput}
\begin{CodeOutput}
[1] "*.1"
\end{CodeOutput}
\begin{CodeInput}
R> node$getChildrenPaths() # Returns the paths of the children nodes
\end{CodeInput}
\begin{CodeOutput}
[1] "*.1.0.0" "*.1.0.1"
\end{CodeOutput}
\begin{CodeInput}
R> node$counts  # The occurrences of symbols following this context
\end{CodeInput}
\begin{CodeOutput}
[1] 13 12
\end{CodeOutput}
\begin{CodeInput}
R> node$extra  # Additional information stored in the node
\end{CodeInput}
\begin{CodeOutput}
$n
[1] 25

$p
[1] 0.52 0.48

$nodeLL
[1] -17.30867
\end{CodeOutput}
\end{CodeChunk}

When a dataset is associated with a \code{ContextTree}, the
\code{counts} attribute of each \code{TreeNode} is automatically updated
to reflect the observed transitions from that context. Additional
quantities are then computed and stored in the \code{extra} attribute as
needed by higher-level algorithms. Because these quantities remain
attached to each node, they can be reused across different inference
procedures without repeated computation. This caching mechanism is one
of the key design features of the package and substantially simplifies
the implementation of algorithms that repeatedly explore different
Context Trees.

The \code{extra} attribute is a modifiable list that can store arbitrary
quantities associated with the node, providing a flexible mechanism for
algorithms that require additional information beyond the default
attributes computed. By default, it contains the following quantities,
which are useful for both frequentist and Bayesian analyses:

\begin{itemize}
  \item \code{n}: The number of times the context represented by the node appears in the data, i.e., 
  $\sum_{k \in \mathcal{A}} n_{s,k}$ in our notation.
  \item \code{p}: The Maximum Likelihood Estimator of the transition probabilities given the context represented by the node. $\hat{p}_{s,k}$ as in \eqref{eq:MLE}.
  \item \code{nodeLL}: The value of the log-likelihood contribution of the node, evaluated at the MLE of the transition probabilities, i.e., $\sum_{k \in \mathcal{A}} n_{s,k} \log(\hat{p}_{s,k})$.
\end{itemize}

When using the \code{baConTree} class, additional Bayesian-specific
quantities are automatically computed and stored in the \code{extra}
attribute. These quantities will be described in Section 3.3, where the
\code{baConTree} class is introduced.

\subsection{ContextTree}\label{contexttree}

While \code{TreeNode} objects represent individual contexts, the
\code{ContextTree} class provides the central abstraction of the package
by organizing these nodes into a complete tree structure and exposing
the operations required to manipulate them.

The key design decision underlying \pkg{bacontrees} is the separation
between the \emph{maximal tree} and the \emph{active tree}. Rather than
representing each candidate Context Tree as an independent object, the
package constructs a single maximal tree containing every context
supported by the data up to a specified maximum depth (\(L\)).
Individual Context Trees are then represented implicitly by selecting
which nodes of this maximal tree are active.

This representation offers several computational advantages. Since all
candidate trees share the same underlying nodes, statistics associated
with a context are computed only once and can be reused across different
models. Structural modifications become local operations that activate
or deactivate nodes while preserving the remaining tree, eliminating the
need to repeatedly construct or destroy tree objects during inference.

The \code{ContextTree} class therefore provides a collection of generic
operations that are independent of any particular statistical
methodology. These include tree traversal (\code{getActiveNodes()},
\code{getInnerNodes()}), node activation and pruning, conversion between
different tree representations, visualization, and validation. As a
result, algorithms with very different statistical objectives can rely
on exactly the same computational infrastructure.

This separation between tree representation and inference algorithms
constitutes one of the main design principles of \pkg{bacontrees}. New
methods can be developed by combining existing tree operations rather
than implementing their own data structures, making the package both
extensible and easier to maintain.

\subsection{baConTree}\label{bacontree}

The \code{baConTree} class extends the general-purpose infrastructure
provided by \code{ContextTree} with methods and quantities required for
Bayesian inference. Importantly, it does not introduce a different tree
representation, but instead augments the existing computational
structure with Bayesian-specific information while preserving the same
underlying nodes and tree organization. The two required new arguments
are the Dirichlet hyperparameter \(\alpha\) and the prior-weight
function \code{priorWeights}, which corresponds to the \(w(\cdot)\)
function in \eqref{eq:bayesian_system}.

Additional quantities stored by the class include prior specifications,
posterior probabilities, marginal likelihood contributions, branching
probabilities, and other values required by Bayesian algorithms. Because
these quantities are attached directly to the nodes, posterior
calculations naturally reuse the same structural representation employed
by other inference procedures.

This design allows Bayesian algorithms such as posterior computation,
MAP estimation, and Markov chain Monte Carlo sampling to operate
directly on the same tree representation used by frequentist procedures.
Consequently, Bayesian and frequentist methods differ primarily in the
statistical quantities they associate with each node rather than in the
underlying data structures used to represent Context Trees.

Another important enhancement when using a \code{baConTree} is that once
the data, the Dirichlet parameter and prior weights are set, several
precomputations are performed and stored in the \code{extra} attribute
of each \code{TreeNode}. These precomputations include the Marginal Tree
Posterior contribution of each node and other quantities needed to
efficiently compute acceptance probabilities during Metropolis--Hastings
sampling and are accessible, allowing them to be used for any other
algorithms. The cached values stored in the \code{extra} attribute of
each node are:

\begin{itemize}
  \item \code{priorWeight}: The prior weight of the node stored as a high-precision \code{brob} number, i.e., $w(s)$.
  \item \code{logPriorWeight}: The log prior weight, i.e., $\log(w(s))$.
  \item \code{nodeLogQ}: The log of the Dirichlet-Multinomial marginal likelihood contribution of the node, i.e., $\log q_{\alpha}(\mathfrak{z})$ from \eqref{eq:marginal_posterior}.
  \item \code{logPosteriorWeight}: The log contribution of the node to the Marginal Tree Posterior from \eqref{eq:marginal_posterior}, i.e., $\log\left(w(s)\, q_{\alpha}(\mathfrak{z})\right)$.
  \item \code{sigmaPrior}, \code{sigmaPosterior}: Recursive partition-function values used for exact posterior computations (e.g., normalizing constants, MAP trees, and sampling).
  \item \code{growPosteriorRatio}, \code{prunePosteriorRatio}: Log acceptance-ratio components for Metropolis–Hastings grow/prune proposals.
\end{itemize}

By extending rather than replacing the \code{ContextTree} class,
\code{baConTree} illustrates the modular philosophy adopted throughout
the package, since statistical methodology is layered on top of a
reusable computational infrastructure instead of being tightly coupled
to a specific implementation.

The package includes a unit test suite built with the \pkg{testthat}
framework \citep{pkgtestthat}, covering all three classes and every
user-facing function. Beyond basic operational checks, the suite
verifies mathematical invariants of the model, such as the consistency
between log and non-log marginal likelihood computations and the
constraint that branching and stopping probabilities must sum to one at
every non-maximal node.

\section{Hands-On Examples}\label{sec:examples}

\subsection{Simulation with rvlmc()}\label{simulation-with-rvlmc}

The \code{rvlmc()} function generates synthetic sequences from a
user-specified VLMC model, defined by a vector of context strings and
the corresponding transition probability vectors:

\begin{CodeChunk}
\begin{CodeInput}
R> alphabet      <- c("a", "b", "c")
R> context_list  <- c("*.a", "*.b", "*.c.a", "*.c.b", "*.c.c")
R> context_probs <- list(
+   c(0.10, 0.20, 0.70),  # *.a
+   c(0.33, 0.33, 0.34),  # *.b
+   c(0.20, 0.10, 0.70),  # *.c.a
+   c(0.01, 0.98, 0.01),  # *.c.b
+   c(0.40, 0.40, 0.20)   # *.c.c
+ )
R> 
R> set.seed(1)
R> sim_seq  <- rvlmc(1000, alphabet, context_list, context_probs)
R> sim_list <- rvlmc(c(500, 500, 500), alphabet, context_list, context_probs)
\end{CodeInput}
\end{CodeChunk}

The function accepts either a scalar \code{n} (returning a single
character vector) or a vector of lengths (returning a list of character
vectors), naturally fitting the data format expected by all other
functions in the package. We use \code{sim\_list} in the following
sections to demonstrate how the estimation methods recover the true
underlying model.

\subsection{Manipulating Context Tree
Structures}\label{manipulating-context-tree-structures}

In this section, we illustrate the basic workflow for creating and
manipulating a \code{ContextTree} object. We begin by loading the
example dataset and building a maximal tree, then demonstrate how to
inspect node properties and navigate between different active-tree
configurations.

\begin{CodeChunk}
\begin{CodeInput}
R> data("abc_list", package = "bacontrees")
R> ct <- ContextTree$new(abc_list, maximalDepth = 3)
R> 
R> # The active tree after initialization is the maximal tree.
R> ct$getActiveNodes()
\end{CodeInput}
\begin{CodeOutput}
 [1] "*.a.b.c" "*.b.c.a" "*.a.a.a" "*.b.c.b" "*.a.a.b" "*.b.c.c" "*.a.a.c"
 [8] "*.b.b.a" "*.b.b.b" "*.b.b.c" "*.c.c.a" "*.b.a.a" "*.c.c.b" "*.b.a.b"
[15] "*.c.b.a" "*.c.c.c" "*.b.a.c" "*.c.b.b" "*.c.b.c" "*.c.a.a" "*.c.a.b"
[22] "*.c.a.c" "*.a.c.a" "*.a.c.b" "*.a.b.a" "*.a.c.c" "*.a.b.b"
\end{CodeOutput}
\end{CodeChunk}

\autoref{fig:ct_example} illustrates how a context tree with a specific
active-tree configuration is represented.

\begin{CodeChunk}
\begin{CodeInput}
R> ct_example <- ContextTree$new(alphabet = c(0,1), maximalDepth = 4)
R> ct_example$activateRoot()
R> ct_example$growActive("*")
R> ct_example$growActive("*.1")
R> ct_example$growActive("*.1.1")
R> ct_example$growActive("*.1.1.0")
R> plot(ct_example, activeOnly = FALSE)
\end{CodeInput}
\begin{figure}[h]

{\centering \includegraphics{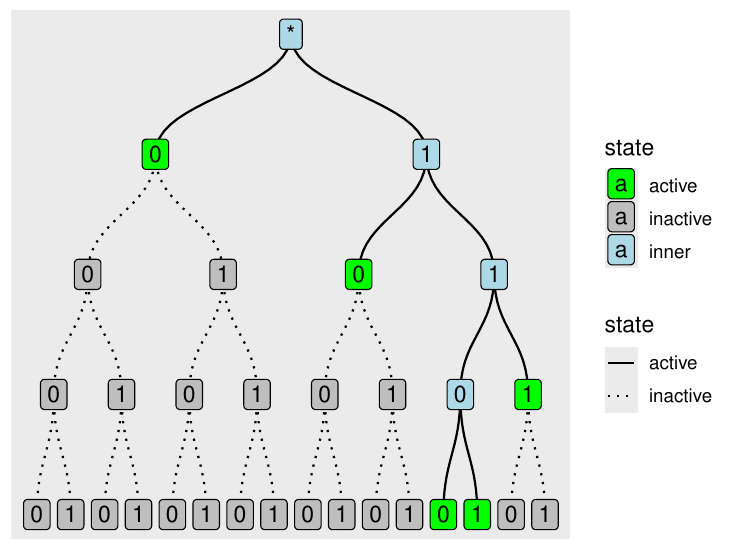} 

}

\caption{Example of active tree representing the same Context Tree as in \autoref{fig:tree_examples}.\label{fig:ct_example}}\label{fig:unnamed-chunk-5}
\end{figure}
\end{CodeChunk}

Active-tree configurations can be set in several ways.
\code{activateRoot()} collapses the tree to a single root node, while
\code{activateMaximal()} restores all leaves of the maximal tree.
Intermediate configurations are obtained by growing or pruning
individual nodes.

\begin{CodeChunk}
\begin{CodeInput}
R> ct$activateRoot()
R> ct$getActiveNodes()   # Only the root is active
\end{CodeInput}
\begin{CodeOutput}
[1] "*"
\end{CodeOutput}
\begin{CodeInput}
R> ct$growActive("*")    # Grow: root becomes inner, its children become active
R> ct$getActiveNodes()   # Now: *.a, *.b, *.c
\end{CodeInput}
\begin{CodeOutput}
[1] "*.a" "*.b" "*.c"
\end{CodeOutput}
\begin{CodeInput}
R> ct$growActive("*.c")  # Grow context *.c
R> ct$getActiveNodes()   # Contexts: *.a, *.b, *.c.a, *.c.b, *.c.c
\end{CodeInput}
\begin{CodeOutput}
[1] "*.a"   "*.c.a" "*.b"   "*.c.b" "*.c.c"
\end{CodeOutput}
\end{CodeChunk}

The current active-tree configuration can be saved as a compact string
code and later restored, which is useful for storing MCMC chains:

\begin{CodeChunk}
\begin{CodeInput}
R> code <- ct$activeTreeCode()
R> code  # A compact binary-encoded string
\end{CodeInput}
\begin{CodeOutput}
[1] "AAAAADNAAA"
\end{CodeOutput}
\begin{CodeInput}
R> ct$activateRoot()          # Change the active tree
R> ct$activateByCode(code)    # Restore from code
R> ct$getActiveNodes()        # Back to the previous configuration
\end{CodeInput}
\begin{CodeOutput}
[1] "*.a"   "*.c.a" "*.b"   "*.c.b" "*.c.c"
\end{CodeOutput}
\end{CodeChunk}

Node-level information is accessed directly from the \code{nodes} list.
When a dataset is associated with the tree, each node stores the
observed transition counts and precomputed statistics in its
\code{extra} field:

\begin{CodeChunk}
\begin{CodeInput}
R> node <- ct$nodes[["*.c"]]
R> node$counts          # Counts of transitions: to a, b, c
\end{CodeInput}
\begin{CodeOutput}
[1] 253 593 400
\end{CodeOutput}
\begin{CodeInput}
R> node$extra$n         # Total observations passing through this context
\end{CodeInput}
\begin{CodeOutput}
[1] 1246
\end{CodeOutput}
\begin{CodeInput}
R> node$extra$nodeLL    # Log-likelihood contribution at the MLE
\end{CodeInput}
\begin{CodeOutput}
[1] -1298.153
\end{CodeOutput}
\end{CodeChunk}

It is also possible to specify an active tree directly from a set of
context strings using \code{activateFromContexts()}, without manually
growing nodes one at a time:

\begin{CodeChunk}
\begin{CodeInput}
R> ct$activateFromContexts(c("*.a", "*.b", "*.c.a", "*.c.b", "*.c.c"))
R> ct$getActiveNodes()
\end{CodeInput}
\begin{CodeOutput}
[1] "*.a"   "*.c.a" "*.b"   "*.c.b" "*.c.c"
\end{CodeOutput}
\end{CodeChunk}

\subsection{Frequentist Estimation}\label{sec:frequentist}

The \code{fit\_vlmc()} function fits a VLMC model to data using the
Context Algorithm \citep{buhlmann1999variable}. It builds the maximal
tree, computes the likelihood-ratio test statistic for each prunable set
of sibling nodes, and prunes those below a user-specified \code{cutoff}.

\begin{CodeChunk}
\begin{CodeInput}
R> fit <- fit_vlmc(sim_list, cutoff = 20, max_length = 3)
R> fit$getActiveNodes()
\end{CodeInput}
\begin{CodeOutput}
[1] "*.a"   "*.c.a" "*.b"   "*.c.b" "*.c.c"
\end{CodeOutput}
\end{CodeChunk}

The \code{cutoff} parameter controls the aggressiveness of pruning:
larger values retain fewer contexts, while smaller values allow longer
memory in the model. For a binary alphabet with one degree of freedom
the \(\chi^2_{1,\, 1-\alpha}\) quantile provides a principled starting
point, as illustrated in \autoref{sec:comparisons}.

\subsection{Bayesian Inference with
baConTree}\label{bayesian-inference-with-bacontree}

For a full Bayesian analysis, users can either use the high-level
\code{metropolis\_vlmc()} wrapper or work directly with the
\code{baConTree} class. We demonstrate both approaches here.

Before running the sampler, we choose \code{maximalDepth} in a
data-driven way using sequential Bayes Factors. For each candidate depth
\(d\), we construct a \code{baConTree} object with a uniform prior
(\(\alpha = 0.5\) and constant node weights) and record its log marginal
likelihood \(\ell_d\). The Bayes Factor
\(\mathrm{BF}(d{+}1, d) = \exp(\ell_{d+1} - \ell_d)\) quantifies the
evidence for extending the tree by one level.

\begin{CodeChunk}
\begin{CodeInput}
R> log_ml <- numeric(5)
R> for (d in 1:5) {
+   bt_d <- baConTree$new(sim_list, maximalDepth = d, alpha = 0.5,
+                         priorWeights = function(node) 1)
+   log_ml[d] <- bt_d$getMarginalLikelihood(log = TRUE)
+ }
R> bf <- exp(diff(log_ml))
R> names(bf) <- paste0("BF(", 2:5, " vs ", 1:4, ")")
R> round(bf, 3)
\end{CodeInput}
\begin{CodeOutput}
  BF(2 vs 1)   BF(3 vs 2)   BF(4 vs 3)   BF(5 vs 4) 
2.047934e+85 1.167000e+00 0.000000e+00 0.000000e+00 
\end{CodeOutput}
\end{CodeChunk}

The Bayes Factor for depth 3 over depth 2 exceeds one, while the factor
for depth 4 over depth 3 falls below one, confirming that
\code{maximalDepth = 3} is well supported by the data.

The \code{metropolis\_vlmc()} function creates a \code{baConTree}
internally, runs the Metropolis-Hastings sampler, and returns the
posterior distribution over context trees as a ranked data frame:

\begin{CodeChunk}
\begin{CodeInput}
R> library(progressr)
R> set.seed(42)
R> with_progress({
+   result <- metropolis_vlmc(
+     sim_list,
+     n_steps         = 2000,
+     max_depth       = 3,
+     alpha           = 0.5,
+     context_weights = function(node) 1,
+     burnin          = 200
+   )
+ })
R> print(result)
\end{CodeInput}
\begin{CodeOutput}
Tree with highest posterior probability ( 0.952 ):
{ *.a, *.c.a, *.b, *.c.b, *.c.c }
# A tibble: 6 x 4
  tree_contexts                                    prob     n tree_code 
  <chr>                                           <dbl> <int> <chr>     
1 {*.a,*.c.a,*.b,*.c.b,*.c.c}                   0.952    1713 AAAAADNAAA
2 {*.c.c.a,*.c.c.b,*.c.c.c,*.a,*.c.a,*.b,*.c.b} 0.0317     57 AACJADMAAA
3 {*.a,*.c.a,*.c.b,*.c.c,*.b.a,*.b.b,*.b.c}     0.00778    14 AAAAADFNAA
4 {*.c.b.a,*.c.b.b,*.c.b.c,*.a,*.c.a,*.b,*.c.c} 0.00444     8 AAACGDJAAA
# i 2 more rows
\end{CodeOutput}
\end{CodeChunk}

The returned object contains a \code{df} data frame with trees ranked by
posterior probability, a \code{codes} list mapping tree codes to context
sets, the full \code{chain} of sampled tree codes, and the underlying
\code{baConTree} object.

Because \code{metropolis\_vlmc()} uses Metropolis-Hastings, the sampled
frequencies are only an approximation to the true posterior. The
included \code{result\$baConTree} gives access to exact posterior
quantities, making it straightforward to compare the MCMC estimate
against the exact posterior probability of the MAP tree:

\begin{CodeChunk}
\begin{CodeInput}
R> # Activate the MAP tree and retrieve its exact posterior probability
R> result$baConTree$activateMap()
R> result$baConTree$getActiveNodes()
\end{CodeInput}
\begin{CodeOutput}
[1] "*.a"   "*.c.a" "*.b"   "*.c.b" "*.c.c"
\end{CodeOutput}
\begin{CodeInput}
R> result$baConTree$activeTreeProbabilities()
\end{CodeInput}
\begin{CodeOutput}
$prior
[1] 0.001369863

$posterior
[1] 0.9572053
\end{CodeOutput}
\end{CodeChunk}

The inner-node structure of sampled trees can also be used to compute
the posterior expected distance to a reference tree.
\citet{paulichen2026context} define the distance between two context
trees as the size of the symmetric difference of their inner node sets,
\(d(\tau_1, \tau_2) = |I(\tau_1) \triangle I(\tau_2)|\), where
\(I(\tau)\) denotes the set of non-leaf nodes of \(\tau\). Using
\code{activateByCode()} and \code{getInnerNodes()}, this distance can be
evaluated for each tree in the MCMC chain returned by
\code{metropolis\_vlmc()}:

\begin{CodeChunk}
\begin{CodeInput}
R> # Inner nodes of the MAP tree (reference)
R> result$baConTree$activateMap()
R> map_inner <- result$baConTree$getInnerNodes()
R> 
R> # Symmetric-difference distance to MAP for each chain tree
R> distances <- sapply(result$chain, function(code) {
+   result$baConTree$activateByCode(code)
+   inner <- result$baConTree$getInnerNodes()
+   length(setdiff(inner, map_inner)) + length(setdiff(map_inner, inner))
+ })
R> mean(distances)
\end{CodeInput}
\begin{CodeOutput}
[1] 0.05247376
\end{CodeOutput}
\end{CodeChunk}

The mean distance summarizes how concentrated the posterior is around
the MAP tree. A value close to zero indicates that most sampled trees
are structurally similar to the MAP tree.

\subsection{Context-Tree Weighting}\label{sec:ctw}

The flexibility of the \code{priorWeights} argument is best illustrated
by one of its most important special cases: the Context-Tree Weighting
(CTW) algorithm \citep{willems2002context}. \citet{paulichen2026context}
show that, for a binary alphabet \(\mathcal{A} = \{0,1\}\) with
hyperparameter \(\alpha = 1/2\), the CTW algorithm corresponds to the
node-weighting prior with weights \begin{equation}\label{eq:ctw-weights}
  w(s) = \begin{cases}
    1/4, & \text{if } s \text{ is an internal node of the maximal tree,} \\
    1/2, & \text{if } s \text{ is a leaf of the maximal tree.}
  \end{cases}
\end{equation} Under this choice the marginal likelihood
\(p(\mathbf{z}; W, \alpha)\) computed by \pkg{bacontrees} equals the
original CTW sequence probability exactly.

The \code{isLeaf()} method of a node object returns \code{TRUE} when the
node sits at the maximum depth of the maximal tree, making it
straightforward to implement the weights in \eqref{eq:ctw-weights}. We
generate a short binary sequence with \code{rvlmc()} and fit a
\code{baConTree} object using those weights.

\begin{CodeChunk}
\begin{CodeInput}
R> # Generate binary data from a known VLMC
R> bin_alphabet <- c("0", "1")
R> bin_contexts <- c("*.0", "*.1.0", "*.1.1")
R> bin_probs    <- list(c(0.8, 0.2), c(0.3, 0.7), c(0.6, 0.4))
R> set.seed(7)
R> bin_seq <- rvlmc(500, bin_alphabet, bin_contexts, bin_probs)
R> 
R> # CTW node-weighting function: 1/2 at maximum depth, 1/4 otherwise
R> ctw_weights <- function(node) if (node$isLeaf()) 0.5 else 0.25
R> 
R> bt_ctw <- baConTree$new(bin_seq, maximalDepth = 4, alpha = 0.5,
+                         priorWeights = ctw_weights)
R> bt_ctw$getMarginalLikelihood(log = TRUE)
\end{CodeInput}
\begin{CodeOutput}
[1] -292.893
\end{CodeOutput}
\end{CodeChunk}

Having seen \pkg{bacontrees} in action, the next section describes the
package's built-in datasets and its visualization system in more detail.

\section{Package Features}\label{sec:features}

\subsection{Example Datasets}\label{example-datasets}

To facilitate testing and demonstration of the package's
functionalities, \pkg{bacontrees} includes two simulated datasets over a
three-symbol alphabet (a, b, c), generated from a known VLMC model with
a specific context tree structure and transition probabilities. These
datasets can be loaded with the \code{data()} function in \proglang{R}
and are available in objects named \code{abc\_vec} and \code{abc\_list}.
\code{abc\_vec} contains a single sequence of 1000 symbols, while
\code{abc\_list} contains a list of 3 independent sequences, each with
1000 symbols.

\begin{CodeChunk}
\begin{CodeInput}
R> data("abc_list", package = "bacontrees")
R> data("abc_vec", package = "bacontrees")
\end{CodeInput}
\end{CodeChunk}

These datasets serve as practical examples for users to explore the
package's capabilities in handling VLMC data and highlight two main
features of the package regarding the type of dataset it can handle:

\begin{itemize}
  \item Both single sequences and multiple independent sequences are supported, accommodating different data structures common in VLMC analyses.
  \item The alphabet is not restricted to numeric values: the datasets use character symbols, and any categorical data can be used directly.
\end{itemize}

\subsubsection{Built-in Functions}\label{built-in-functions}

While the core strength of \pkg{bacontrees} lies in its flexible data
structures and methods for tree manipulation, it also includes a suite
of built-in functions that implement common workflows for
variable-length Markov chains and Bayesian context-tree models. These
functions leverage the underlying classes to provide high-level
operations that are frequently needed in practice. This allows users to
perform standard tasks without having to implement them from scratch,
while still benefiting from the modular and extensible design of the
package. This makes the package accessible to users who need
ready-to-use implementations without needing to interact with the
underlying classes directly.

Some of the key built-in functions include:

\begin{itemize}
  \item \code{rvlmc()}: Generates synthetic sequences of specified lengths, from a specified VLMC model defined by a context tree and transition probabilities.
  \item \code{metropolis\_vlmc()}: Performs Metropolis-Hastings sampling for VLMC models and produces results in a comprehensive format, including the posterior distribution over trees as a ranked data frame.
  \item \code{fit\_vlmc()}: Fits a VLMC model to observed data using the Context Algorithm from \citet{buhlmann1999variable}, pruning the maximal tree via likelihood-ratio tests.
\end{itemize}

\subsection{Tree Visualization}\label{tree-visualization}

\pkg{bacontrees} provides specialized plotting functions for visualizing
context tree structures, leveraging the capabilities of the \pkg{ggraph}
package \citep{pkgggraph}. These functions allow users to create clear
and informative visual representations of context trees, which can be
essential for understanding their structure and properties.

By default, the \code{plot()} functions for \code{ContextTree} and
\code{baConTree} objects generate a basic visualization of the tree
structure, displaying nodes and edges in a hierarchical layout. These
plot functions are built on top of \pkg{ggraph}, thus returning a
\code{ggraph} object, allowing users to further customize the plots by
adding layers or modifying aesthetics using standard \pkg{ggraph} and
\pkg{ggplot2} \citep{pkgggplot2} syntax. Examples of how to customize
the plots are provided below in \autoref{fig:customplot1} and
\autoref{fig:customplot2}.

For \code{ContextTree} objects, users can choose to visualize either the
active tree or the full maximal tree by setting the \code{activeOnly}
parameter. The state of each node (active or inactive) is visually
distinguished using different colors or styles. If data is associated
with the tree, edges have their widths scaled according to the counts of
transitions between contexts, providing additional insight into the
observed sequences.

\begin{CodeChunk}
\begin{CodeInput}
R> library(ggplot2)
R> ct <- ContextTree$new(abc_list, maximalDepth = 2)
R> ct$activateRoot()
R> ct$growActive("*")
R> ct$growActive("*.c")
R> ct$growActive("*.b")
R> plot(ct, activeOnly = FALSE) + 
+   ggtitle("Edge Widths Proportional to Counts")
\end{CodeInput}
\begin{figure}[h]

{\centering \includegraphics{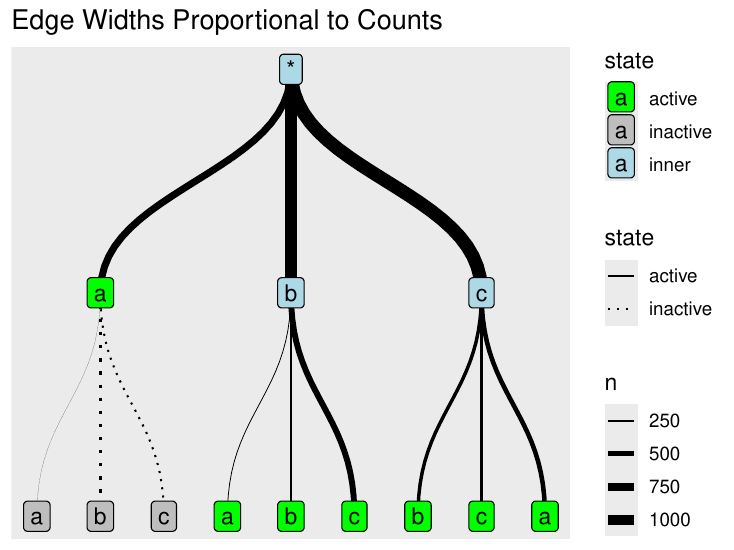} 

}

\caption{Plot of a maximal Context Tree with a custom title and edge widths proportional to observed counts.\label{fig:customplot1}}\label{fig:unnamed-chunk-17}
\end{figure}
\end{CodeChunk}

\begin{CodeChunk}
\begin{CodeInput}
R> plot(ct, activeOnly = TRUE) +
+   theme_void() + theme(legend.position = "none") +
+   scale_fill_manual(values = c("active" = "orange", "inner" = "white"))
\end{CodeInput}
\begin{figure}[h]

{\centering \includegraphics[width=0.25\linewidth]{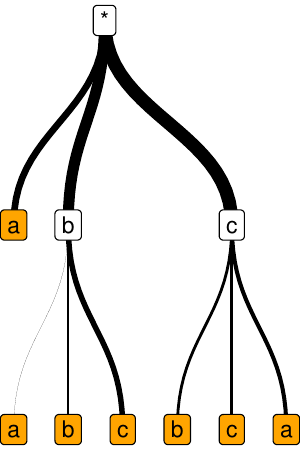} 

}

\caption{Plot of an active Context Tree without any additional elements.\label{fig:customplot2}}\label{fig:unnamed-chunk-18}
\end{figure}
\end{CodeChunk}

For \code{baConTree} objects, the maximal tree is always visualized,
with nodes colored according to the log-ratio of having that node as a
context versus pruning it (together with its siblings) and replacing
them with the parent node. A simple interpretation is that negative
values indicate that pruning the node (and its siblings together) would
decrease the posterior probability of the tree, making the node more
likely to be part of the true context tree, whereas values close to zero
or positive suggest the node contributes little and is more likely to be
pruned. \autoref{fig:customplot3} shows the default plot for a
\code{baConTree} object. For the dataset used in this example,
\code{abc\_list}, the true context tree has 5 contexts: \(\{*.a \}\),
\(\{*.b \}\), \(\{*.c.a\}\), \(\{*.c.b\}\) and \(\{*.c.c\}\). The plot
indicates that the nodes corresponding to these contexts have negative
posterior log-ratios, suggesting they are more likely to be part of the
true context tree, while other nodes have log-ratios close to 0,
indicating they are less likely to be included.

\begin{CodeChunk}
\begin{CodeInput}
R> bct <- baConTree$new(abc_list, maximalDepth = 3, alpha = 0.5,
+                      priorWeights = function(node) 1)  # Uniform prior weights
R> plot(bct)
\end{CodeInput}
\begin{figure}[h]

{\centering \includegraphics{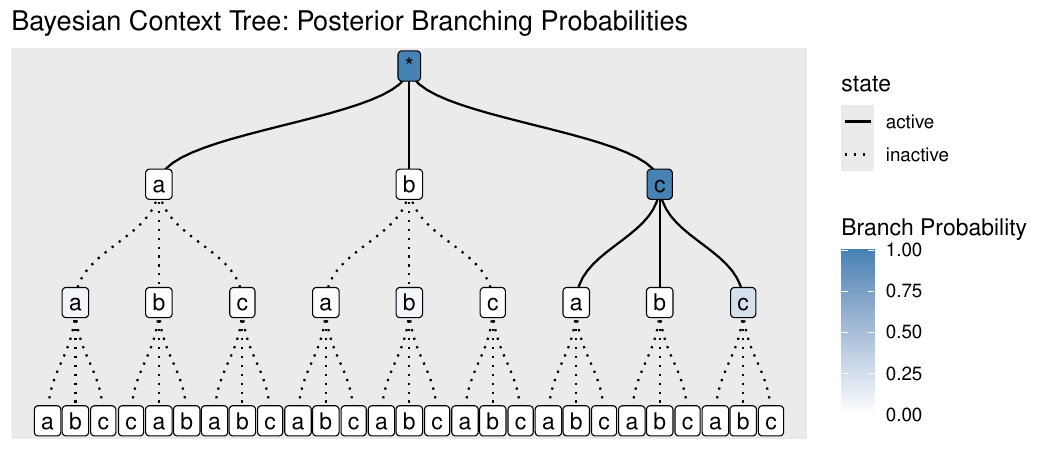} 

}

\caption{Plot of a \code{baConTree} with nodes colored according to posterior weights.\label{fig:customplot3}}\label{fig:unnamed-chunk-19}
\end{figure}
\end{CodeChunk}

\section{Comparisons to other packages}\label{sec:comparisons}

Two other \proglang{R} packages address related problems and are worth
comparing with \pkg{bacontrees}: \pkg{VLMC} \citep{pkgVLMC}, which
provides the frequentist Context Algorithm, and \pkg{BCT}
\citep{pkgbct}, which implements the Bayesian Context Trees framework of
\citet{kontoyiannis2022bayesian}. \autoref{tab:package_comparison}
summarizes their respective feature sets.

\begin{table}[ht]
\centering
\begin{tabular}{r|ccc}
\toprule
Feature & \pkg{bacontrees} & \pkg{VLMC} & \pkg{BCT} \\ \midrule
Context Algorithm (frequentist VLMC) & $\checkmark$ & $\checkmark$ & $\times$ \\
Bayesian posterior sampling (MCMC)   & $\checkmark$ & $\times$     & $\checkmark$ \\
Exact MAP tree inference             & $\checkmark$ & $\times$     & $\checkmark$ \\
Marginal likelihood                  & $\checkmark$ & $\times$     & $\checkmark$ \\
Multiple independent sequences       & $\checkmark$ & $\times$     & $\times$ \\
Custom prior weights                 & $\checkmark$ & $\times$     & $\times$ \\
Exposed tree node structure          & $\checkmark$ & $\times$     & $\times$ \\
Simulation utilities                 & $\checkmark$ & $\checkmark$ & $\checkmark$ \\
Plotting capabilities                & $\checkmark$ & $\checkmark$ & $\checkmark$ \\
Extensible / customizable plots      & $\checkmark$ & $\times$     & $\times$ \\
\pkg{igraph} interoperability        & $\checkmark$ & $\times$     & $\times$ \\
\bottomrule
\end{tabular}
\caption{Comparison of features across context-tree packages for \proglang{R}.
\label{tab:package_comparison}}
\end{table}

\subsection{Frequentist Estimation with
VLMC}\label{frequentist-estimation-with-vlmc}

The \pkg{VLMC} package \citep{pkgVLMC} provides the classical Context
Algorithm for fitting variable-length Markov chains by significance
testing. It accepts character, numeric, or factor vectors and prunes the
maximal tree using a chi-squared threshold controlled by the
\code{alpha.c} argument. We fit a model to the binary sequence
\code{bin\_seq} defined in \autoref{sec:ctw}:

\begin{CodeChunk}
\begin{CodeInput}
R> library("VLMC")
R> # bin_seq is a character vector "0"/"1" defined in Section 4.5
R> vlmc_fit <- vlmc(bin_seq, alpha.c = 0.05)
R> summary(vlmc_fit)
\end{CodeInput}
\begin{CodeOutput}

'vlmc' a Variable Length Markov Chain;
     alphabet '01', |alphabet| = 2, n = 500.
Call: vlmc(dts = bin_seq, alpha.c = 0.05);  cutoff =1.920729
 -> extensions (= $size ) :
   ord.MC   context nr.leaves     total 
        4         7         2        34 
AIC =  562.1555 
R^2 = %{correctly predicted} = 72.6%
Confusion matrix:
    predicted
data   0   1
   0 313  19
   1 117  50
Markov chain depths along the data:
    Min.  1st Qu.   Median     Mean  3rd Qu.     Max. 
       1        1        2 1.873747        2        4 
\end{CodeOutput}
\end{CodeChunk}

The tree can be displayed as an ASCII dendrogram with \code{draw()}:

\begin{CodeChunk}
\begin{CodeInput}
R> draw(vlmc_fit)
\end{CodeInput}
\begin{CodeOutput}
[x]-( 332 167| 499)-F
 +--[0]-( 259 73| 332) <10.56>
 |   '--[0]-( 208 51| 259) <0.41>
 |       '--[1]-( 48 3| 51) <3.99>-T
 '--[1]-( 73 94| 167) <18.21>
     '--[0]-( 22 50| 72) <2.62>
         '--[0]-( 18 33| 51) <0.26>
             '--[1]-( 3 0| 3) <3.12>-T
\end{CodeOutput}
\end{CodeChunk}

The corresponding \pkg{bacontrees} fit uses \code{fit\_vlmc()} with a
chi-squared cutoff. For a binary alphabet the likelihood-ratio test has
one degree of freedom, so \(\alpha = 0.05\) corresponds to a threshold
of \(\chi^2_{1,\,0.95} \approx 3.84\):

\begin{CodeChunk}
\begin{CodeInput}
R> fit_bin <- fit_vlmc(list(bin_seq), cutoff = qchisq(0.95, 1), max_length = 4)
R> fit_bin$getActiveNodes()
\end{CodeInput}
\begin{CodeOutput}
[1] "*.0.1"   "*.1.0"   "*.1.1"   "*.0.0.0" "*.0.0.1"
\end{CodeOutput}
\end{CodeChunk}

The two fits need not identify exactly the same context tree, for two
reasons. First, \pkg{bacontrees} considers only complete context trees,
where sibling nodes are always pruned or kept together, whereas
\pkg{VLMC} can retain a single suffix without its sibling. Second,
\pkg{bacontrees} excludes the first \(D\) observations of each sequence
from the likelihood (where \(D\) is the \code{maximalDepth} parameter),
while \pkg{VLMC} uses the full sequence, which changes the evidence on
which the pruning decision is based. The two tests serve the same
purpose of selecting the context tree, and users can adjust the cutoff
to obtain comparable results, keeping these differences in mind. The
\pkg{VLMC} package is well-suited for frequentist model selection and
sequence prediction \citep{pkgVLMC}, but it does not expose the internal
tree structure, making it difficult to extend or inspect the fitted
model beyond the functions the package provides. \pkg{bacontrees}
additionally returns live node objects that can be traversed, annotated,
or fed into further algorithms (see \autoref{sec:frequentist}), and it
supports multiple independent sequences.

\subsection{Bayesian Inference with
BCT}\label{bayesian-inference-with-bct}

The \pkg{BCT} package \citep{pkgbct} implements the Bayesian Context
Trees framework of \citet{kontoyiannis2022bayesian}, with functions for
computing the MAP context tree, the marginal likelihood (via the CTW
algorithm), and top-\(k\) model rankings. Its input must be a single
collapsed character string, rather than a vector.

\begin{CodeChunk}
\begin{CodeInput}
R> library("BCT")
R> # BCT requires the sequence as one collapsed string
R> bin_str <- paste(bin_seq, collapse = "")
R> 
R> # MAP context tree at maximum depth 4
R> bct_map <- BCT(bin_str, depth = 4)
R> bct_map$Contexts   # leaf context strings of the MAP tree
\end{CodeInput}
\begin{CodeOutput}
[1] "0"    "11"   "101"  "1000" "1001"
\end{CodeOutput}
\end{CodeChunk}

The log marginal likelihood under the CTW prior is returned by
\code{CTW()}:

\begin{CodeChunk}
\begin{CodeInput}
R> CTW(bin_str, depth = 4)
\end{CodeInput}
\begin{CodeOutput}
[1] -292.893
\end{CodeOutput}
\end{CodeChunk}

For comparison, the same quantity computed by \pkg{bacontrees} with the
CTW node-weighting function (see \autoref{sec:ctw}) yields:

\begin{CodeChunk}
\begin{CodeInput}
R> bt_ctw$getMarginalLikelihood(log = TRUE)
\end{CodeInput}
\begin{CodeOutput}
[1] -292.893
\end{CodeOutput}
\end{CodeChunk}

The two values agree. The MAP context tree returned by \pkg{BCT} can be
compared with the one obtained from \pkg{bacontrees} by activating the
MAP tree of the fitted \code{baConTree} object:

\begin{CodeChunk}
\begin{CodeInput}
R> bt_ctw$activateMap()
R> bt_ctw$getActiveNodes()
\end{CodeInput}
\begin{CodeOutput}
[1] "*.1.1"     "*.0"       "*.1.0.1"   "*.1.0.0.0" "*.1.0.0.1"
\end{CodeOutput}
\end{CodeChunk}

The active nodes of \code{bt\_ctw} after \code{activateMap()} correspond
to the same contexts reported in \code{bct\_map\$Contexts}, confirming
that both packages recover the same MAP tree. Unlike \pkg{BCT},
\pkg{bacontrees} allows the prior weights to be replaced by any
user-defined function, supports multiple independent sequences, and
gives direct access to the tree's node objects for custom traversals or
diagnostics.

\subsection{Positioning of bacontrees}\label{positioning-of-bacontrees}

\pkg{bacontrees} is designed to complement both packages rather than
replace them. For pure frequentist use, \pkg{VLMC} offers a mature and
well-tested implementation. For binary sequences under the fixed BCT
prior, \pkg{BCT} is a direct and efficient choice. \pkg{bacontrees}
targets users who need greater flexibility: the ability to specify
custom prior distributions, work with non-binary or multi-sequence data,
navigate the tree programmatically, or build new algorithms on top of
the shared data structures.

General-purpose packages for hierarchical data structures, such as
\pkg{data.tree} \citep{pkgdata.tree}, \pkg{igraph} \citep{pkgigraph},
and \pkg{tidygraph} \citep{pkgtidygraph}, provide powerful abstractions
for generic trees and graphs. However, they are intentionally agnostic
about the statistical properties that characterize context trees (node
counts, alphabet-specific likelihoods, context activation logic, and the
posterior quantities needed for Bayesian inference), so substantial
additional implementation work would be needed before any of them could
support context-tree algorithms directly. \pkg{bacontrees} fills this
gap by providing a self-contained framework where the statistical
semantics are built into the data structure itself. With its position
among existing tools clarified, the next section offers a brief summary
of the package's contributions.

\section{Conclusions}\label{sec:conclusions}

The \pkg{bacontrees} package provides a unified and extensible framework
for representing, manipulating, and performing inference with context
trees in \proglang{R}. Its design is centered on three \pkg{R6} classes,
\code{TreeNode}, \code{ContextTree}, and \code{baConTree}, which
encapsulate the core data structures and algorithms needed for both
frequentist and Bayesian analyses of discrete sequential data.

The package covers the full workflow from data generation and model
fitting to visualization and posterior inference. Simulation utilities,
exact MAP tree computation, marginal likelihood evaluation compatible
with the Context-Tree Weighting algorithm, and Metropolis-Hastings
posterior sampling are all available through a consistent interface. The
layered class hierarchy allows users to implement custom inference
algorithms without reimplementing the underlying tree operations, and
the \pkg{ggplot2}-based visualization system supports
publication-quality figures with minimal additional code.

The package is available on CRAN as well as on GitHub
(\url{https://github.com/Freguglia/bacontrees}), where development
versions are hosted and issues can be reported.

\section*{Acknowledgments}\label{acknowledgments}
\addcontentsline{toc}{section}{Acknowledgments}

This work was supported by FAEPEX (Fundo de Apoio ao Ensino, à Pesquisa
e à Extensão), Universidade Estadual de Campinas (Unicamp), under grant
number 3387/23.

\vfill
\pagebreak

\bibliography{refs.bib}

@article{freguglia2023detecting,
  title={Detecting renewal states in chains of variable length via intrinsic Bayes factors},
  author={Freguglia, Victor and Garcia, Nancy L},
  journal={Statistics and Computing},
  volume={33},
  number={1},
  pages={1--17},
  year={2023},
  publisher={Springer}
}

@article{buhlmann1999variable,
  title={Variable length Markov chains},
  author={B{\"u}hlmann, Peter and Wyner, Abraham J},
  journal={The Annals of Statistics},
  volume={27},
  number={2},
  pages={480--513},
  year={1999},
  publisher={Institute of Mathematical Statistics}
}

@article{willems2002context,
  title={The context-tree weighting method: Basic properties},
  author={Willems, Frans MJ and Shtarkov, Yuri M and Tjalkens, Tjalling J},
  journal={IEEE transactions on information theory},
  volume={41},
  number={3},
  pages={653--664},
  year={2002},
  publisher={IEEE}
}

@article{kontoyiannis2022bayesian,
  title={Bayesian context trees: Modelling and exact inference for discrete time series},
  author={Kontoyiannis, Ioannis and Mertzanis, Lambros and Panotopoulou, Athina and Papageorgiou, Ioannis and Skoularidou, Maria},
  journal={Journal of the Royal Statistical Society Series B: Statistical Methodology},
  volume={84},
  number={4},
  pages={1287--1323},
  year={2022},
  publisher={Oxford University Press}
}

@Manual{pkgbct,
  title = {BCT: Bayesian Context Trees for Discrete Time Series},
  author = {Ioannis Papageorgiou and Valentinian Mihai Lungu and Ioannis Kontoyiannis},
  year = {2022},
  note = {R package version 1.2},
  url = {https://CRAN.R-project.org/package=BCT},
  doi = {10.32614/CRAN.package.BCT},
}

@Manual{pkgVLMC,
  title = {VLMC: Variable Length Markov Chains ('VLMC') Models},
  author = {Martin Maechler},
  year = {2024},
  note = {R package version 1.4-4},
  url = {https://CRAN.R-project.org/package=VLMC},
  doi = {10.32614/CRAN.package.VLMC},
}

@Manual{pkgigraph,
  title = {{igraph}: Network Analysis and Visualization in R},
  author = {Gábor Csárdi and Tamás Nepusz and Vincent Traag and Szabolcs Horvát and Fabio Zanini and Daniel Noom and Kirill Müller and David Schoch and Maëlle Salmon},
  year = {2025},
  note = {R package version 2.2.1},
  doi = {10.5281/zenodo.7682609},
  url = {https://CRAN.R-project.org/package=igraph},
}

@Manual{pkgdata.tree,
  title = {data.tree: General Purpose Hierarchical Data Structure},
  author = {Christoph Glur},
  year = {2025},
  note = {R package version 1.2.0},
  url = {https://CRAN.R-project.org/package=data.tree},
  doi = {10.32614/CRAN.package.data.tree},
}

@Manual{pkgggraph,
  title = {ggraph: An Implementation of Grammar of Graphics for Graphs and Networks},
  author = {Thomas Lin Pedersen},
  year = {2025},
  note = {R package version 2.2.2},
  url = {https://CRAN.R-project.org/package=ggraph},
  doi = {10.32614/CRAN.package.ggraph},
}

@Book{pkgggplot2,
  author = {Hadley Wickham},
  title = {ggplot2: Elegant Graphics for Data Analysis},
  publisher = {Springer-Verlag New York},
  year = {2016},
  isbn = {978-3-319-24277-4},
  url = {https://ggplot2.tidyverse.org},
}

@Manual{pkgtidygraph,
  title = {tidygraph: A Tidy API for Graph Manipulation},
  author = {Thomas Lin Pedersen},
  year = {2024},
  note = {R package version 1.3.1},
  url = {https://CRAN.R-project.org/package=tidygraph},
  doi = {10.32614/CRAN.package.tidygraph},
}

@article{krichevsky1981performance,
  title={The performance of universal encoding},
  author={Krichevsky, Raphail and Trofimov, Victor},
  journal={IEEE Transactions on Information Theory},
  volume={27},
  number={2},
  pages={199--207},
  year={1981},
  publisher={IEEE}
}

@article{paulichen2026context,
  title={Context Tree Prior Distributions based on Node Weighting with exact Bayes Factors},
  author={Paulichen, Thiago and Freguglia, Victor},
  journal={arXiv preprint arXiv:2603.25806},
  year={2026}
}

@Article{pkgtestthat,
  author  = {Hadley Wickham},
  title   = {testthat: Get Started with Testing},
  journal = {The R Journal},
  year    = {2011},
  volume  = {3},
  number  = {1},
  pages   = {5--10},
  doi     = {10.32614/RJ-2011-002},
}

\end{document}